\documentclass[amsmath,floatfix,twocolumn,superscriptaddress,prx,longbibliography]{revtex4-1}
\usepackage{graphicx,array}
\usepackage[utf8]{inputenc}
\usepackage[english]{babel}

\renewcommand{\vr}{\vec{r}}
\newcommand{\va}{\vec{a}}
\newcommand{\vc}{\vec{c}}

\newcommand{\half}{\frac12}
\newcommand{\DFA}{\text{DFA}}
\newcommand{\Hrm}{\text{H}}

\usepackage[normalem]{ulem}
\usepackage{color}
\usepackage{xcolor}

\definecolor{Mygrey}{gray}{0.80}
\definecolor{orange}{rgb}{1,0.5,0}

\newcommand*{\LightComments}{}%

\ifdefined\NoComments

\else
 \ifdefined\LightComments

 \else

 \fi
\fi

\begin{document}
\title{Combining density functional theories to correctly describe the energy, lattice structure and electronic density of functional oxide perovskites}

\author{Kiel T. Williams} \affiliation{Department of Physics, University of Illinois 
at Urbana-Champaign, Urbana, Illinois 61801, USA}

\author{Lucas K. Wagner} \affiliation{Department of Physics, University of Illinois 
at Urbana-Champaign, Urbana, Illinois 61801, USA}

\author{Claudio Cazorla} \affiliation{School of Materials Science and
  Engineering, UNSW Australia, Sydney NSW 2052, Australia}

\author{Tim Gould} \affiliation{Queensland Micro- and Nanotechnology
  Centre, Griffith University, Nathan, QLD 4111, Australia}

\maketitle

{\bf Functional oxide perovskites are the pillar of cutting-edge technological applications. Density functional theory (DFT) simulations are the theoretical methods of choice to understand and design perovskite materials. However, tests on the reliability of DFT to describe fundamental properties of oxide perovskites are scarce and mostly ill-defined due to a lack of rigorous theoretical benchmarks for solids. Here, we present a quantum Monte Carlo benchmark study of DFT on the archetypal perovskite BaTiO$_{3}$ (BTO). It shows that no DFT approximation can simultaneously reproduce the energy, structure, and electronic density of BTO. Traditional protocols to select DFT approximations are empirical and fail to detect this shortcoming. An approach combining two different non-empirical DFT schemes, ``SCAN''~\cite{scan} and ``HSE06''~\cite{hse06}, is able to holistically describe BTO with accuracy. Combined DFT approaches should thus be considered as a promising alternative to standard methods for simulating oxide perovskites.
}
\\

Following the pioneering works by Vanderbilt~\cite{vanderbilt93,vanderbilt94} and Cohen~\cite{cohen92}, density functional theory~\cite{KohnSham} (DFT) has emerged as the preferred quantum method for simulating and understanding functional oxide perovskites with chemical formula $AB$O$_{3}$. Oxide perovskites are prototypical ferroelectric, multiferroic, piezoelectric, and flexoelectric materials that find numerous applications in sensors, transducers, memory devices and energy converters, to cite just a few examples~\cite{spaldin19,rappe13,cazorla19,catalan16}. The great functionality of oxide perovskites stems from the ability to induce abrupt structural and electronic changes via modest thermodynamic shifts and external fields.

\begin{figure}[t]
  \includegraphics[width=\linewidth]{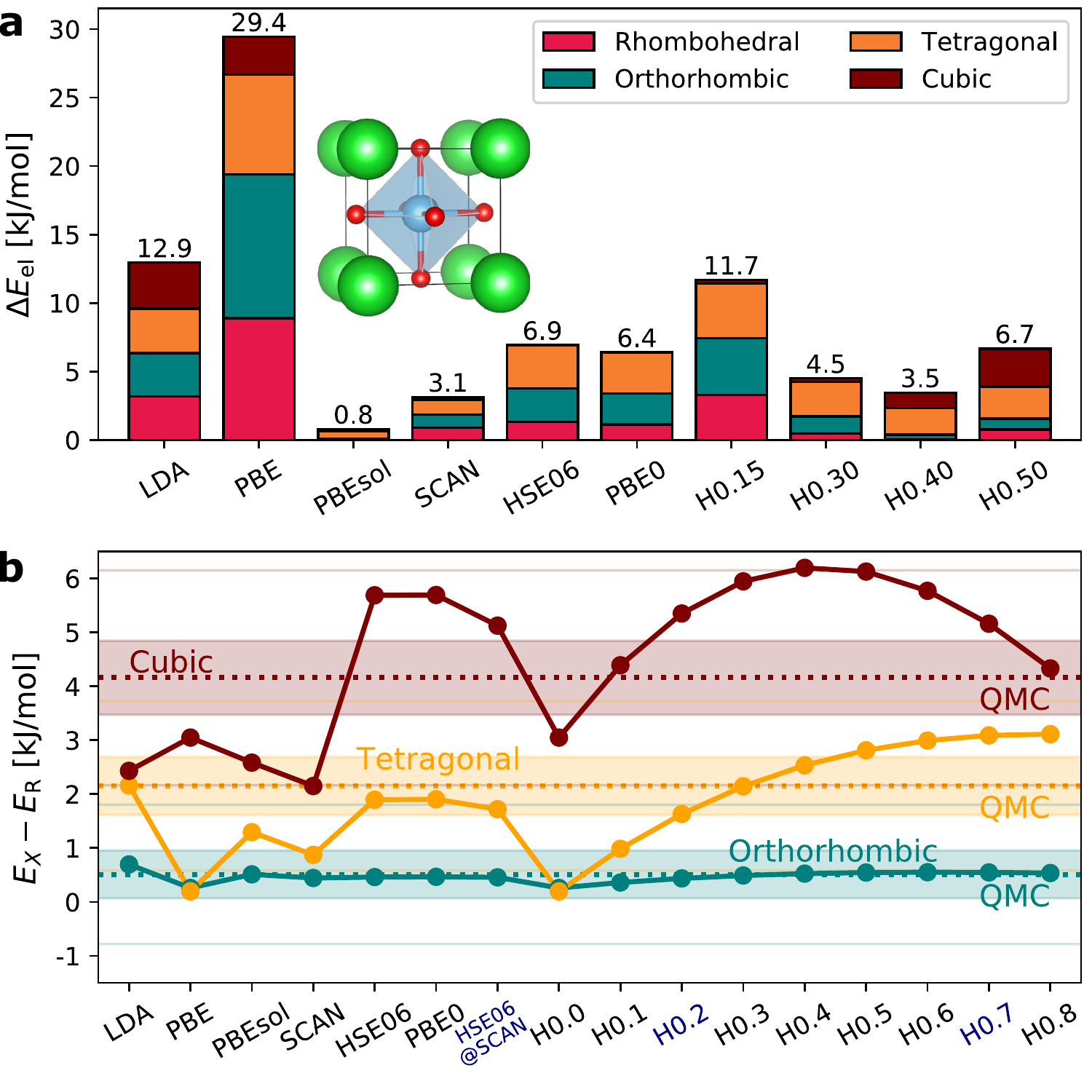}
  \caption{{\bf a}~Elastic energy metric [Eq.\eqref{eqn:EEl}] of lattice structural errors for a variety of DFA applied to four BTO polymorphs. The best approaches (PBEsol, SCAN) have $\Delta E_{\text{el}}$ close to zero. Colours indicate the contribution from different polymorphs, which are summed fo the overall ranking. The inset shows the cubic structure. {\bf b}~Energy differences relative to the R structure for a longer list of DFA compared against benchmark QMC results. The experimentally reported structures~\cite{edwards51,cheong93} are considered in all the cases. The shaded regions indicate confidence interval  for QMC energies. The best approaches (H0.2, H0.7 and HSE06@SCAN) have energies for all four polymorphs close to the confidence intervals.}
\label{fig:Energies}
\end{figure}

Accurate first-principles simulation of functional oxide perovskites is challenging. Firstly, the presence of strongly correlated $d$ electrons introduced by the transition metal (TM) $B$ ions leads to severe self-interaction and charge delocalisation DFT errors~\cite{franchini14}. Secondly, the delicate balance between short-range and long-range forces~\cite{cohen92}, originated by strong TM $d$ and O $p$ electronic orbital hybridizations and the Coulomb interactions between permanent electric dipoles and higher order moments, is difficult to reproduce with an unique DFT approximation (DFA). Thirdly, the energy differences between crystal polymorphs typically are tiny ($\sim 1$~kJ~mol$^{-1}$) and thus genuinely accurate computational methods are needed to predict realistic phase transformations. Importantly, these ``hard-to-model'' features are inherent to the extraordinary functionality that characterizes oxide perovskites.

The traditional approach to modelling oxide perovskites is empirical, using a set of experimental data, typically lattice parameters and energy band gaps, to select the DFA that best reproduces it. To improve the agreement with experiments while simultaneously limiting the computational burden, more often than not \emph{ad hoc} energy terms are added to the constitutive DFT Kohn-Sham equations (e.g., DFT+$U$~\cite{dudarev98}). The above-described protocol has two key limitations, even in the idealistic case of error-free experiments: (i) reproducing correct lattice parameters does not imply exactness of other quantities; (ii) there is no theoretical justification for good band gaps even in exact DFT~\cite{Perdew2017}. Furthermore, experimental comparisons present the additional handicap that energy differences between polymorphs are neglected.

To rigorously determine the performance of a computational method it is necessary to gauge its results against benchmarks obtained with higher accuracy methods and under identical modelling conditions. In quantum chemistry, benchmarking is customarily done for molecules and other non-extended systems by using advanced ``wavefunction theories'' (WFT)~\cite{benchmark1,benchmark2} that take advantage of the boundness of molecular wavefunctions in real space. In materials science, similar benchmark tests are very scarce due to the technical complexity of applying WFT to periodic systems. Hybrid DFT functionals are commonly regarded as the gold standard for oxide perovskites because they can partially cure self-interaction errors (SIE)~\cite{bilc2008,evarestov12}. However, it remains unclear how well hybrid DFA work for relevant quantities like electronic densities and energy differences~\cite{cazorla17,Williams2020}.

This work presents a benchmark diffusion quantum Monte Carlo (QMC) study which assesses the performance of common DFA in describing the archetypal oxide perovskite BaTiO$_{3}$ (BTO), both for the energy and density. QMC is an accurate many-body wavefunction method that only suffers from a relatively small (compared to DFA) fixed node approximation~\cite{wagner16,wagner15,wagner07,foulkes01}. Benchmarked DFAs are the local density approximation~\cite{lda} (LDA), two generalized gradient approximations (GGA: PBE~\cite{pbe}, PBEsol~\cite{pbesol}), one meta-GGA (SCAN~\cite{scan}), one range-separated hybrid (HSE06~\cite{hse06}) and various hybrids (PBE0~\cite{pbe0}, H$c$). The ``H$c$'' functionals are of PBE0 form~\cite{pbe0}, but with an arbitrary fraction of exact exchange, $c$, namely: $E_{\text{xc}}^{\Hrm{c}}=cE_{\text{x}}^{\text{HF}} + (1-c)E_{\text{x}}^{\text{PBE}} + E_{\text{c}}^{\text{PBE}}$. The $c=0.25$ functional, H0.25, is equivalent to PBE0. Dispersion corrections were also tested but found to have poor performance and are thus discussed in the Supplementary Material.

We begin our discussion with the traditional analysis of comparing the theoretically predicted lattice parameters with the available experimental data. BTO has four polymorphs (rhombohedral, orthorhombic, tetragonal and cubic -- R, O, T and C, respectively) which are all nearly degenerate in energy and have very similar structures~\cite{cazorla19}. To allow for a direct ranking of DFA, we introduce a physically motivated metric to convert errors in lattice parameters into a single number: the total elastic strain energy per functional unit, $\Delta E_{\text{el}}^{\DFA} =\sum_{X\in\{\rm R,O,T,C\}}\Delta E_{\text{el},X}^{\DFA}$, summed over the four polymorph strain energies (Methods),
\begin{align}
  \Delta E_{\text{el},X}^{\DFA}
	=&\half V_{X}\sum_{kl}D_{kl,X}\Delta_{k,X}^{\DFA}\Delta_{l,X}^{\DFA}\;.
  \label{eqn:EEl}
\end{align}
Parameters $D_{kl,X}$ and $V_{X}$ are obtained from PBEsol calculations (Supplementary Material). $\Delta_{k,X}^{\DFA}$, defined as $(a_{k,X}^{\DFA}/a_{k,X}^{\text{expt}}) - 1$, indicates the error of the lattice parameter $a_{k,X}$ computed by DFA relative to its experimental value for the same structure $X$~\cite{edwards51,cheong93}. Equation~\eqref{eqn:EEl} thus distinguishes errors of little importance from those which are great. The results shown in Fig.\ref{fig:Energies}a point to PBEsol as the clear winner since the corresponding structures are almost in perfect agreement with the experiments (Supplementary Material). SCAN is a little worse, but still very good. By tradition, PBEsol would be a recommended DFA to continue modelling BTO.

Our ultimate goal is to understand and simulate the functionality of oxide perovskites. Hence, in addition to predicting accurate structures we also need to estimate accurate energy differences between polymorphs. This information usually is not available because measurement of internal energies, $E_X$, is not possible from experiments and application of WFT methods to solid-state systems is complex and arduous. As mentioned above, here we make up for such a lack of benchmarks by carrying out QMC calculations (Methods).

The DFT energy differences, $E_X-E_R$, for the four BTO polymorphs relative to the ground-state phase (R) are shown in Fig.\ref{fig:Energies}b, where they are compared against the QMC results. The experimentally reported structures~\cite{edwards51,cheong93} are considered in all the cases. It is readily apparent from the figure that PBEsol is not, in fact, very good at energetics, despite its outstanding success at predicting lattice structures. The traditional approach to choosing a DFA is thus quite questionable. Indeed no approach is perfectly accurate in predicting energy differences between all polymorphs, although a few (H0.2, H0.7 and HSE06@SCAN -- the latter will be introduced later) are noticeably precise.

\begin{figure}[t]
  \includegraphics[width=\linewidth]{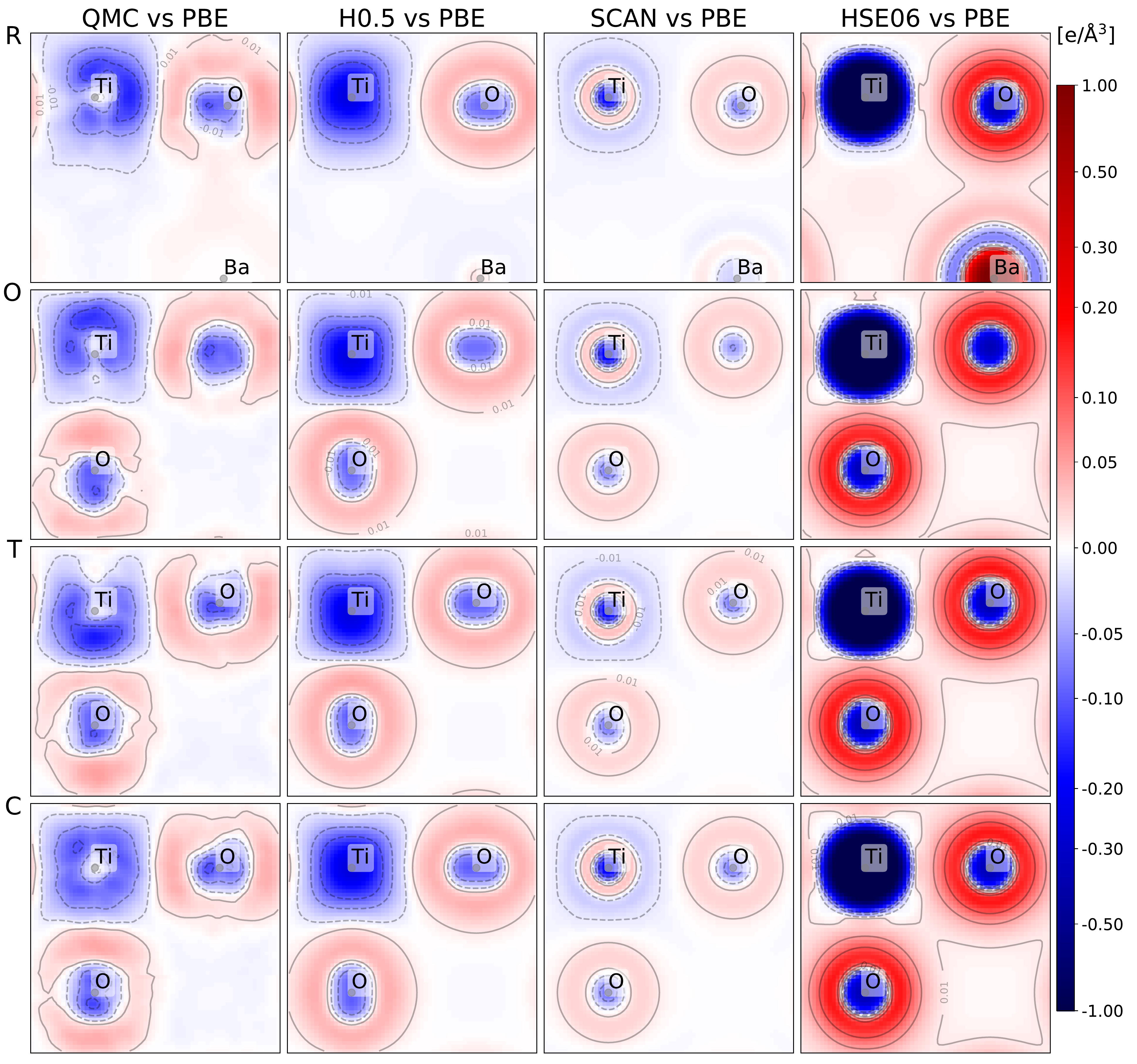}
  \caption{Electronic density differences for BTO polymorphs on a Ti--O plane, shown for QMC and selected DFA relative to PBE densities. The far left column (QMC vs PBE) represents the target. Deviations from this target in other columns  indicate errors in the densities. The best approach is H0.5. A non-linear scale has been employed due to the need to represent exponentially varying data.}
\label{fig:Densities}
\end{figure}

Density functional theory relies on the electronic density as its primary variable.  Thus, we next seek to determine which DFA gives a density closest to the QMC benchmark. We cannot meaningfully quantify this difference (Methods)~\cite{Medvedev2017} thus instead rely on qualitative visual comparisons. Figure~\ref{fig:Densities} shows density differences between DFA and PBE (both calculated deterministically) against differences between QMC and PBE (both computed stochastically). For all structures, we consider the density on a plane containing one Ti cation and at least one O atom. The distribution of electrons in this plane is very relevant because metal-oxide physics suffer notoriously from SIE and we thus expect density errors to be greatest therein.  For compactness, we restrict our analysis to a representative subset of DFA: PBE, SCAN and HSE06 and H0.5 (additional tests for other DFA and perovskite planes can be found in the Supplementary Material).

Two notable points stand out from Fig.\ref{fig:Densities}. Firstly, H0.5 gives good densities, especially in the critical Ti--O bonding region. The H0.5 oxygen shell structure approximately matches that of QMC, and both present similarities along the bonds. The main differences between QMC and H0.5 are found in the Ti nucleus (where numerical differences are expected, Methods), and in the circular (QMC) versus slightly elliptical (H0.5) O densities. Nonetheless, the H0.5 density seems to be a very good substitute for the QMC density, a feature that we shall use later. Both approaches transfer more charge from Ti (more blue) to O (more red) ions than PBE does. Secondly, the popular HSE06 functional (range-separated hybrid) gives quite bad densities, with significantly larger amounts of charge transferred from Ti to O ions than QMC. This shortcoming occurs despite the fact that HSE06 provides some of the best energy differences between polymorphs. On the other hand, it seems consistent with the mediocre performance of HSE06 in predicting structures (Fig.\ref{fig:Energies}a). We also see that SCAN gets qualitatively good features but has a slight excess (depletion) of charge on the O (Ti) ions as compared to QMC. PBEsol turns out to be equally poor than PBE on the electronic densities despite its outstanding success in predicting lattice parameters (Supplementary Material).

\begin{figure}[t]
   \includegraphics[width=\linewidth]{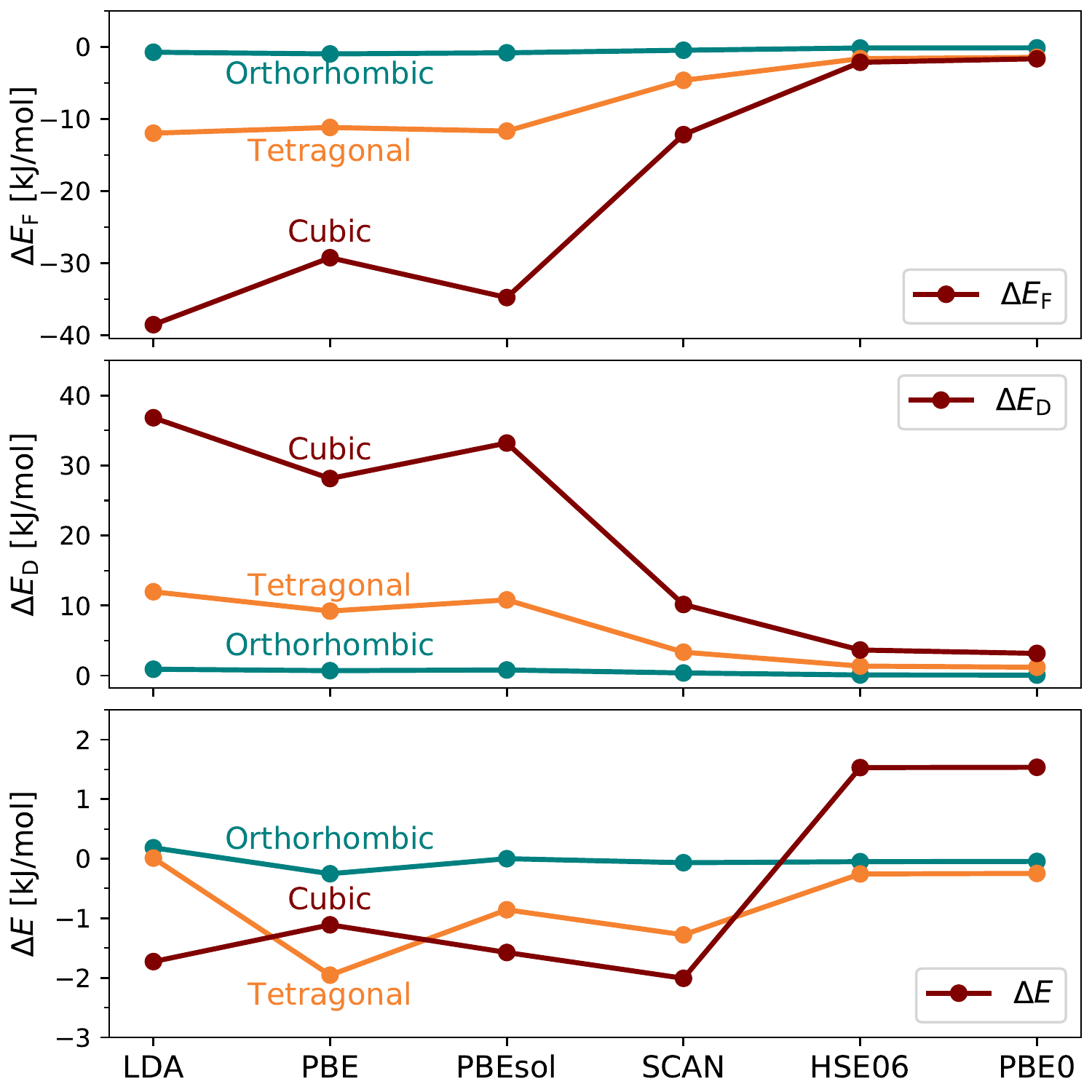}
   \caption{Functional-driven ($\Delta E_{\text{F}}$, top panel), density-driven ($\Delta E_{\text{D}}$, middle) and total ($\Delta E=\Delta E_{\text{F}}+\Delta E_{\text{D}}$, bottom -- note different scale) errors in the energies for various DFA (relative to R). All functionals benefit from cancellation of errors, with larger values of $\Delta E_{\text{F}/\text{D}}$ likely to lead to larger non-systematic $\Delta E$ errors.}
\label{fig:DD}
\end{figure}

To complete our analysis, we seek to assess the relevance of these density errors and thus, hopefully, offer strategies toward higher quality calculations. We turn to a recent work by Kim, Sim and Burke~\cite{kim13}, who proposed a physically motivated approach for splitting the error of DFA into useful components, called functional-driven and density-driven errors. These metrics quantify the importance of SIE on a given problem and thus provide an estimate of DFA performance.

The QMC density cannot, due to numerical noise, be used directly to estimate functional-driven and density-driven errors. However, H0.5 is a good approximation to the QMC density (Fig.~\ref{fig:Densities}). Thus, denoting $n_{\Hrm{0.5}}$ the H0.5 density, $n_{\DFA}$ the self-consistent density of a given DFA, and $E^{\DFA}[n]$ the energy obtained using DFA on an arbitrary density $n$, we use
\begin{align}
  \Delta E^{\DFA}_{\text{F}} =& E^{\DFA}[n_{\Hrm{0.5}}] - E^{\text{QMC}}\;,
  \\
  \Delta E^{\DFA}_{\text{D}} =& E^{\DFA}[n_{\DFA}] - E^{\DFA}[n_{\Hrm{0.5}}]\;,
\end{align}
for the energies associated with errors from the approximate functional $E^{\DFA}$ and approximate density $n_{\DFA}$, respectively. Then, the total energy error is $\Delta E^{\DFA} = \Delta E^{\DFA}_{\text{F}} + \Delta E^{\DFA}_{\text{D}}$ (Fig.\ref{fig:DD}). We see that LDA, PBE and PBEsol all benefit from enormous cancellation of density and functional errors in the T and C polymorphs, with the individual components being up to 10 times as large as the total error. By contrast, PBE0 and HSE06 always have density and functional errors of similar scale to their total errors. Thus, they are less sensitive to density errors and tend to get trends correctly. Nonetheless, the agreement with the QMC benchmarks is worsened in all the cases by using H0.5 densities (compare $\Delta E_{\text{F}}$ with $\Delta E$).

We have shown that no single functional describes correctly the general properties of BTO, a model oxide perovskite.  Most DFA fail to reproduce all three metrics reported here (structure, energy and density) while a few succeed in one. Thus, any DFA is guaranteed to be compromised. This is not surprising of oxide perovskites, which are highly prone to SIE and where the lattice and electronic degrees of freedom are strongly coupled~\cite{vanderbilt94,cohen92,spaldin19}. A more comprehensive search over the growing range of DFA would, with no doubt, find an approach which does succeed in mimicking BTO. But, it seems unlikely that such an approach would succeed equally well in describing other metal--oxide systems since DFA achievements mostly rely on large and non-systematic cancellations of errors (Fig.\ref{fig:DD}).

\begin{figure}[t]
  \includegraphics[width=\linewidth]{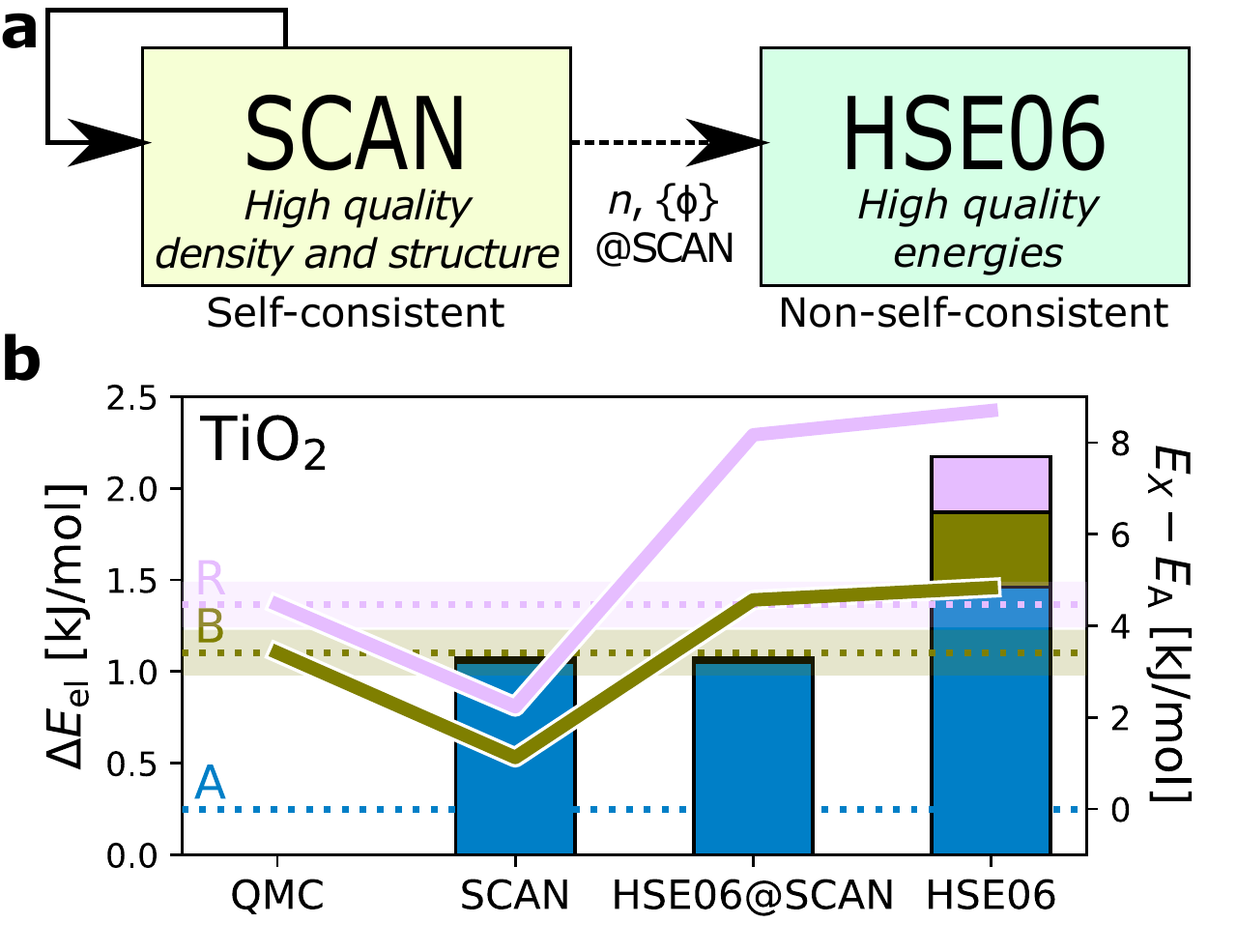}
  \caption{{\bf a} The HSE06@SCAN method uses self-consistent SCAN calculations of structures and densities and non-self-consistent HSE06 calculations of energies, to achieve holistic success in predicting electronic structure. {\bf b} Elastic strain energy (rectangles) and energy differences relative to the ground state (lines) calculated for the TiO$_{2}$ polymorphs anatase (A), brookite (B) and rutile (R). HSE06@SCAN out-performs HSE06 on the structures and energies and SCAN on the energies. The confidence interval for the QMC energies~\cite{kent15} are indicated with shaded regions.}
\label{fig:HSE06SCAN}
\end{figure}

Thus, rather than seeking a \emph{single DFA} that can succeed in general, we recommend an alternative strategy for modelling functional oxide perovskites: \emph{merging two DFA to combine the best of both}. For this purpose, we propose carrying out HSE06 (range-separated hybrid) energy calculations using the density, $n$, and orbitals, $\{\phi\}$, from a converged and structurally optimized SCAN (meta-generalized gradient approximation) calculation, which we denote here as
HSE06@SCAN and illustrate in Fig.\ref{fig:HSE06SCAN}a.

As shown in Fig.\ref{fig:Energies}b, HSE06@SCAN provides accurate relative energies for all BTO polymorphs (slightly more accurate than HSE06, which is already one of the best performers) while simultaneously benefiting from the correct lattice parameters and electronic densities obtained with SCAN (Figs.\ref{fig:Energies}a and \ref{fig:Densities}). To further justify this recommendation, we note that (i)~both SCAN and HSE06 are physically well-justified approaches that are constructed to be versatile, and (ii)~HSE06 is relatively insensitive to density-driven errors (Fig.\ref{fig:DD}). Finally, we performed an additional HSE06@SCAN benchmark test for TiO$_{2}$ using QMC data from previous work~\cite{kent15} (Fig.~\ref{fig:HSE06SCAN}b), which shows that HSE06@SCAN improves on both HSE06 (energies and structure) and SCAN (energies). Despite the fact that TiO$_{2}$ is a binary oxide, these supplementary findings further illustrate the efficacy of the proposed HSE06@SCAN  approach.

Merging two cost-effective and non-empirical DFA to improve the overall description of functional oxide perovskites is a new and persuasive recommendation for the broad community of computational materials scientists. Additional testing is certainly required to establish the best combination of DFA for a general solution. Nonetheless, we expect that our work will promote better practices and the appreciation of QMC benchmark studies in the context of functional perovskites modelling.
\\

%

\section*{Methods}
{\bf Quantum Monte Carlo.}~Diffusion quantum Monte Carlo (DMC) calculations were performed using the QWalk software package~\cite{wagner_qwalk:_2009}.  A single determinant composed of PBE0 Kohn-Sham orbitals was generated using the CRYSTAL software package~\cite{dovesi_quantum-mechanical_2018}. The single determinant was multiplied by a 2-body Jastrow factor, optimized with respect to the total energy, according to the linear method~\cite{umrigar_alleviation_2007}.  The resulting wavefunction was used as the diffusion Monte Carlo guiding function.  The $T$-moves scheme was used to ensure a rigorous upper-bound to the true ground state energy~\cite{casula_beyond_2006}. The DMC timestep was extrapolated to zero using a linear fitting procedure and calculations performed at timesteps of $\tau=0.01$ and $\tau=0.02$ Ha$^{-1}$. All DMC quantities were calculated at each point of a $2 \times 2 \times 2$ ${\bf k}$-point grid, and averaged for a simulation cell containing 8 formula units of BTO.  Each DMC calculation was performed using the experimental geometry of the corresponding BTO structural phase. Comparisons between the electronic density obtained with mixed-estimator DMC and the corresponding extrapolated-estimator density show that the mixed-estimator error for these systems is smaller than the Monte Carlo stochastic error. The Burkatzki-Filippi-Dolg pseudopotentials were used to represent ionic cores~\cite{burkatzki_energy-consistent_2007}.
 \\

{\bf Density Functional Theory.}~DFT calculations were performed with the Vienna {\em ab initio} software package (VASP)~\cite{vasp}. The ``projected augmented wave'' (PAW) method~\cite{paw} was employed to represent the ionic cores, and the following electronic states were considered as valence: Ti $3d^{2}4s^{2}3p^{6}$, Ba $6s^{2}5p^{6}5s^{2}$, and O $2p^{4}2s^{2}$. An energy cut-off of $850$~eV and a $\Gamma$-centered ${\bf k}$-point grid of $12 \times 12 \times 12$ were employed in the calculations for a simulation cell containing $5$ atoms. Geometry relaxations were performed with a conjugated gradient algorithm that allowed for changes in the cell volume, cell shape and atomic positions. The geometry relaxations were stopped once the forces acting on the ions were below $0.005$~eV$\cdot$\AA$^{-1}$. Additional details of our DFT calculations can be found in the Supplementary Material.
\\

{\bf Structures.}~Any deformation from the relaxed structure of a crystal has an associated elastic energy cost.  To leading order, this is quadratic with respect to the deformations, and thus can be used as a simple metric for errors in lattice parameters. In the primitive cell, we can define three lattice vectors $\va_{i}$ and assume rigid deformations along these vectors that scale them to $(1+\epsilon_{i})$ of their relaxed length. To simplify matters, shear contributions are ignored. The accompanying elastic energy then is $\Delta E_{\text{el}}=\half V \sum_{ij}\overline{C}_{ij} \epsilon_i\epsilon_j$. where $V$ converts from energy per unit volume to energy per functional unit. In some cases, the crystallographic axes consist of linear combinations of the set $\va_i$, that is, $\vc_i=\sum_ju_{ij}\va_j$, thus giving $\va_i=\sum_jq_{ij}\vc_j$. Scaling these $\vc$ axes by $(1+\Delta_i)$ then gives $\Delta E_{\text{el}}=\half V \sum_{ijkl}q_{ki}q_{lj}\overline{C}_{ij}\Delta_k\Delta_l$, where $D_{kl}=\sum_{kl}q_{ki}q_{lj}\overline{C}_{ij}$ in Eq.\eqref{eqn:EEl}. For our purposes, it is sufficient to directly calculate $D_{ij}$ and $V$ using PBEsol as it gave the most accurate lattice constants. \\

{\bf Densities.}~A simple metric for comparing densities is to take the mean absolute error $\overline{\Delta n}=V_{\text{cell}}^{-1} \int_{\text{cell}}|n_{\DFA}(\vr)-n_{\text{QMC}}(\vr)|d\vr$ (e.g., in Ref.~\cite{Medvedev2017}). However, QMC calculations are stochastic which makes it difficult to distinguish true errors from density fluctuations caused by importance sampling. Some of these issues can be alleviated by comparing differences QMC-PBE calculated stochastically against DFA-PBE calculated deterministically, as we have done in Fig.\ref{fig:Densities}. An additional complication is that the DFT calculations used PAW pseudopotentials with 38 electrons, whereas the QMC calculations used regular pseudopotentials with 32 electrons (6 less on the Ti atoms). This leads to systematic differences between the two sets of results near the Ti nucleus. To further support the results of the manuscript we include many additional density plots in the supplementary material; and comparisons of PBE0 and PBE using the DFT and QMC codes.
\\

\bibliography{bto-qmcdft}

\section*{Data availability}
The data that support the findings of this study are available from the corresponding authors upon reasonable request.\\

\section*{Acknowledgments}
 Computational resources and technical assistance were provided by the Australian Government and the Government of Western Australia through Magnus under the National Computational Merit Allocation Scheme (ca11, ge7) and The Pawsey Supercomputing Centre. K.T.W. and L.K.W. were supported by a grant from the Simons Foundation as part of the Simons Collaboration on the many-electron problem.

\section*{Author contributions}
C.C. and T.G. designed the research. T.G. performed DFT energy calculations. C.C. carried out DFT energy and structural calculations. K.T.W. and L.K.W. performed the QMC calculations. T.G. carried out the  analysis of the electronic density differences. All the authors participated in the writing of the manuscript.\\

\section*{Additional information}
Supplementary information is available in the online version of the paper.\\

\section*{Competing financial interests}
The authors declare no competing financial interests.\\

\end{document}